\documentclass[%
  reprint,
  superscriptaddress,
  nofootinbib,
  amsmath, amssymb,
  aps,
  prc,
  floatfix,
  nolongbibliography,
]{revtex4-2}

\usepackage{graphicx}
\usepackage{color}
\usepackage[breaklinks,colorlinks,urlcolor=blue,citecolor=blue,linkcolor=blue]{hyperref}
\usepackage{palatino}
\usepackage{mathpazo}
\usepackage{xspace}
\usepackage{orcidlink}
\usepackage{bm}
\usepackage{physics}

\graphicspath{{figures/}}

\newcommand{\TransThrustAxis}{\ifmmode{\hat{n}_{\perp}}\else\mbox{$\hat{n}_{\perp}$}\xspace\fi}
\newcommand{\TransThrust}{\ifmmode{\tau_{\perp}}\else\mbox{$\tau_{\perp}$}\xspace\fi}
\newcommand{\Sperp}{\ifmmode{S_{\perp}}\else\mbox{$S_{\perp}$}\xspace\fi}
\newcommand{\eCM}[2]{\ifmmode{\sqrt{s}=#1\,\mathrm{#2}}%
  \else
  \mbox{$\sqrt{s}=#1\,\mathrm{#2}$}\xspace
  \fi
}
\newcommand{\eCMNN}[2]{\ifmmode{\sqrt{s_{\mathrm{NN}}}=#1\,\mathrm{#2}}%
  \else
  \mbox{$\sqrt{s_{\mathrm{NN}}}=#1\,\mathrm{#2}$}\xspace
  \fi
}
\newcommand{\pT}{\ifmmode{p_t}\else\mbox{$p_t$}\xspace\fi}
\newcommand{\pTjet}{\ifmmode{p_t^{\mathrm{jet}}}\else\mbox{$p_t^{\mathrm{jet}}$}\xspace\fi}

\newcommand{\Jewel}{\textsc{Jewel}\xspace}
\newcommand{\FastJet}{\textsc{FastJet}\xspace}
\newcommand{\GeV}{\ifmmode{\,\mathrm{GeV}}\else{\,GeV}\xspace\fi}
\newcommand{\pp}{\ifmmode{p\text{+}p}\else\mbox{$p$+$p$}\xspace\fi}
\newcommand{\pbpb}{\ifmmode{\mathrm{Pb}\text{+}\mathrm{Pb}}%
  \else
  \mbox{$\mathrm{Pb}$+$\mathrm{Pb}$}\xspace
  \fi
}

\usepackage{CJK}

\begin{document}

\begin{CJK*}{UTF8}{gbsn}
\title{Novel observables for exploring jet quenching effects in individual jets}

\author{Jin-Wen Kang (康锦文)\,\orcidlink{0000-0002-4153-0505}}
\affiliation{%
  Key Laboratory of Quark \& Lepton Physics (MOE) and Institute of Particle Physics,
  Central China Normal University, Wuhan 430079, China
}

\author{Sa Wang (王洒)\,\orcidlink{0000-0001-6372-7352}}
\affiliation{%
  Center for Astronomy and Space Sciences and Institute of Modern Physics, China Three Gorges University, Yichang 443002, China
}

\author{Meng-Quan Yang (杨梦权)}
\affiliation{%
  State Key Laboratory of Nuclear Physics and Technology, Institute of Quantum Matter,
  South China Normal University, Guangzhou 510006, China
}
\affiliation{%
  Guangdong Basic Research Center of Excellence for Structure and Fundamental Interactions of Matter,
  Guangdong Provincial Key Laboratory of Nuclear Science, Guangzhou 510006, China
}%

\author{Ben-Wei Zhang (张本威)\,\orcidlink{0000-0001-6075-6609}}
\email[Corresponding author: ]{bwzhang@mail.ccnu.edu.cn}
\affiliation{%
  Key Laboratory of Quark \& Lepton Physics (MOE) and Institute of Particle Physics,
  Central China Normal University, Wuhan 430079, China
}%

\author{Enke Wang (王恩科)}
\affiliation{%
  State Key Laboratory of Nuclear Physics and Technology, Institute of Quantum Matter,
  South China Normal University, Guangzhou 510006, China
}
\affiliation{%
  Guangdong Basic Research Center of Excellence for Structure and Fundamental Interactions of Matter,
  Guangdong Provincial Key Laboratory of Nuclear Science, Guangzhou 510006, China
}%

\date{\today}

\begin{abstract}
  We introduce event-shape observables in the jet frame to characterize intra-jet shapes and study medium modifications in heavy-ion collisions.
  Using the \textsc{Jewel} model, we investigate jet constituent angular and transverse momentum distributions, together with two representative event-shape observables, transverse thrust $\tau_\perp$ and transverse sphericity $S_\perp$, in $p$+$p$ and 0--10\% central Pb+Pb collisions at $\sqrt{s_{\mathrm{NN}}} = 5.02\,\mathrm{TeV}$.
  These observables are evaluated after transforming the jet constituent momenta into the jet frame.
  The intra-jet shape distributions in Pb+Pb collisions shift toward larger values relative to those in $p$+$p$ collisions, indicating more isotropic distributions of constituent transverse momenta.
  Comparisons between calculations with and without recoil, along with the dependence on the constituent transverse momentum threshold, indicate that medium response plays an important role in these modifications.
  These results suggest that jet-frame event-shape observables provide sensitive probes of medium response in jet quenching, complementary to single-particle angular and transverse momentum distributions.
\end{abstract}

\maketitle

\end{CJK*}

\section{Introduction}

In high-energy collisions, jets are collimated sprays of particles that arise when high-\pT quarks or gluons produced in hard scattering processes undergo parton showering and hadronization.
Jets generated at an early stage of the heavy-ion collisions traverse the hot and dense quark-gluon plasma (QGP) medium~\cite{Busza:2018rrf,Zhang:2021xib}, leading to energy loss and modifications of their internal structure; this phenomenon is known as jet quenching~\cite{Wang:1994fx,Gyulassy:2000er,Arnold:2001ms,Baier:1996kr,Zakharov:1997uu,Wiedemann:2000za,Guo:2000nz,Zhang:2003wk,Mehtar-Tani:2025rty,Vitev:2009rd,Qin:2015srf,Cao:2024pxc,Connors:2017ptx,Marzani:2019hun}.
Therefore, jets serve as effective probes for studying the properties of the QGP and the mechanisms of parton energy loss in the medium.
After decades of experimental and theoretical efforts, significant progress has been made in understanding jet quenching in heavy-ion collisions at the Large Hadron Collider (LHC) and the Relativistic Heavy Ion Collider (RHIC).
Specifically, various jet observables have been developed to investigate the formation and properties of the QGP using jet quenching effects as a probe, including jet yields, angular correlations and momentum imbalance in dijet and boson-jet systems, and numerous jet substructure observables, etc.
Relevant measurements have improved our understanding of jet--medium interactions and the underlying mechanisms of jet quenching~\cite{Connors:2017ptx,Marzani:2019hun,ATLAS:2018gwx,Wang:2021jgm,He:2011pd,Li:2024pfi,Li:2024uzk,Dai:2012am,Kang:2023ycg,Wang:2024plm,Zhang:2018urd,ATLAS:2018dgb,ALICE:2019ykw,CMS:2018fof,STAR:2023pal,Caucal:2021cfb,Wang:2019xey,Luo:2018pto,Wang:2025mkw}.
However, the energy distribution of particles within jets and its modifications by the QGP medium remain under active investigation and are not yet fully understood. One of the earliest observables used to study the intra-jet shape is the jet shape function~\cite{CDF:1992cus,Vitev:2008rz,Luo:2018pto,CMS:2018jco}, which only measures the radial distribution of transverse momentum within a jet.
This raises the question of whether we can develop methods to quantify the geometrical patterns of energy flow inside jets.

Event-shape variables are a class of observables that characterize the geometrical properties of energy flow in the final state of high-energy particles collisions, and have been measured in $e^+e^-$ collisions~\cite{ALEPH:2003obs,Dasgupta:2003iq}, $ep$ collisions~\cite{ZEUS:2006vwm}, $p\bar{p}$ collisions~\cite{CDF:1991etg} and \pp collisions~\cite{CMS:2014tkl,CMS:2025sws,ATLAS:2012uka,ALICE:2012cor,ATLAS:2012tch,CMS:2018svp,ATLAS:2020vup}.
Recent theoretical studies have begun to explore event-shape variables in heavy-ion collisions, and in particular have investigated the nuclear modification of event-shape variables in multijet final states in heavy-ion collisions~\cite{Chen:2020pfa,Kang:2023qxb,Kong:2024wdk}. In these studies, the authors found that three event-shape variables tend to be less isotropically distributed in \pbpb collisions than in \pp collisions.
This trend is attributed to the reduction of jet multiplicity under a given selection criterion due to energy loss of jets traversing the QGP, which is contrary to the theoretical expectation that transverse momentum broadening and medium-induced gluon radiation would lead to a more isotropic state.

In this work, we propose to extend the application of event-shape variables to the study of particle distribution inside jets, i.e., the intra-jet shape.
By defining event-shape variables in the jet frame, we aim to capture the geometrical patterns of energy flow within individual jets. This will provide a new perspective for understanding the jet quenching effect on the intra-jet structure.
We will use the \Jewel model to investigate the distributions of these jet-frame event-shape variables in \pp collisions and their modifications due to jet--medium interactions in \pbpb collisions.

This paper is organized as follows. In Sec.~\ref{sec:jet-frame-evtshape}, we introduce the event-shape variables defined in the jet frame. In Sec.~\ref{sec:framework}, we describe the theoretical framework used in this work. Then, we present our numerical results and discussion for the intra-jet shape observables in Sec.~\ref{sec:results}.
Finally, a summary is given in Sec.~\ref{sec:summary}.

\section{Event-shape variables in the jet frame}
\label{sec:jet-frame-evtshape}

Event-shape variables can be used to describe the patterns, correlations, and origins of the energy flow in multijet or hadronic final states~\cite{ATLAS:2012tch}. We present the first calculation of event-shape variables defined in the jet frame for individual jets, which can be used to characterize the intra-jet shape.

\subsection{Event-shape definitions}
\label{subsec:evt-shape-def}

\begin{figure*}
  \centering
  \includegraphics[width=0.45\linewidth]{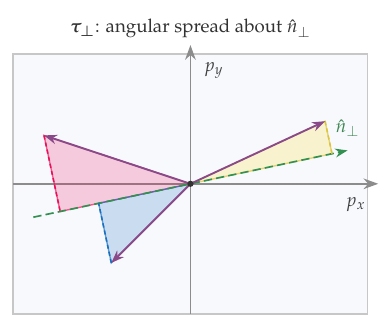}
  \hskip10pt
  \includegraphics[width=0.45\linewidth]{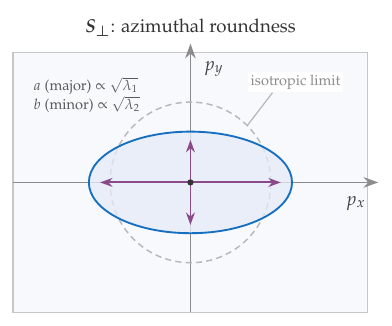}
  \caption{Geometric interpretation of the transverse event-shape observables
    \TransThrust\ and \Sperp. Left: The transverse-momentum vectors are projected onto the transverse thrust axis \TransThrustAxis. The colored triangles illustrate the corresponding projections. The observable \TransThrust\ provides a momentum-weighted measure of the angular spread about \TransThrustAxis\ and approaches zero in the collimated, back-to-back limit.
    Right: The observable $S_\perp$ characterizes the azimuthal roundness of the momentum
    distribution through the eigenvalues $\lambda_1\geq\lambda_2$ of its normalized
    second-moment tensor. The principal semi-axes satisfy
    $a\propto\sqrt{\lambda_1}$ and $b\propto\sqrt{\lambda_2}$, while the dashed
    circle indicates the isotropic limit. Accordingly,
    $S_\perp=2\lambda_2/(\lambda_1+\lambda_2)$ varies from zero for a pencil-like
  distribution to unity for an isotropic distribution.}
  \label{fig:event-shape-illustration}
\end{figure*}

Event-shape variables comprise a broad class of observables designed to characterize the geometry of the energy-momentum flow in the final-state. Many commonly used event-shape variables fall into two classical families: thrust-family and sphericity-family observables.

The thrust-family event-shape observables include thrust, thrust major, thrust minor, transverse thrust, and transverse thrust minor. The (transverse) thrust axis, is also used to define hemisphere-based observables such as total and wide broadening, as well as total and heavy jet masses.
The transverse thrust axis \TransThrustAxis\ is defined as the unit vector in the transverse plane that maximizes the quantity $T_{\perp}$~\cite{ATLAS:2012tch,Banfi:2010xy,Cesarotti:2020hwb,Cesarotti:2020ngq,CMS:2014tkl,ATLAS:2020vup,Prasad:2025yfj}:
\begin{equation}
  T_{\perp}=\max_{\TransThrustAxis} \frac{\sum_i |\vec{p}_{t,i} \cdot \TransThrustAxis|}{\sum_i p_{t,i}},
  \label{equ:Tperp-def}
\end{equation}
where the sum runs over all objects (particles or jets) in the region of investigation, and $\vec{p}_{t,i}$ is the transverse momentum vector of object $i$. The transverse thrust variable \TransThrust\ is then defined as $\TransThrust = 1 - T_{\perp}$.
As illustrated in the left panel of Fig.~\ref{fig:event-shape-illustration}, \TransThrust\ quantifies the momentum-weighted angular spread about the transverse thrust axis, \TransThrustAxis. In other words, \TransThrust\ measures the degree to which the transverse-momentum flow is collimated along the transverse thrust axis. A smaller value of \TransThrust\ indicates a more collimated configuration (i.e., \textit{cigar-like}), whereas a larger value indicates a more isotropic configuration. For a perfectly balanced back-to-back configuration in the transverse plane, \TransThrust\ vanishes. By contrast, for an isotropic configuration, it reaches its theoretical maximum value of $1-2/\pi$.

The sphericity-family event-shape observables include sphericity, aplanarity, planarity, $F$-parameter, and transverse sphericity. These observables are defined in terms of the ordered eigenvalues of the momentum tensor, also referred to as the sphericty tensor.
There are three definitions of the sphericity tensor, quadratic~\cite{Counihan:1975rr,ATHENS-LIVERPOOL-VIENNA:1977fca,Okorokov:2011mb,ALEPH:2003obs,Banfi:2010xy,CMS:2025sws,ATLAS:2012uka}, linearized~\cite{ATLAS:2012uka,ALICE:2012cor,Parisi:1978eg,Donoghue:1979vi,Prasad:2025yfj,ATLAS:2020vup} and generalized~\cite{Cesarotti:2020ngq}.
The quadratic sphericity tensor is not collinear safe~\cite{ALICE:2012cor}. Therefore, in this work we adopt the linearized sphericity tensor, which is both infrared and collinear safe.
The linearized sphericity tensor $\mathcal{M}_{xyz}$ is defined as
\begin{equation}
  \mathcal{M}_{xyz} = \frac{1}{\sum_i |\vec{p}_i|}\sum_i \frac{1}{|\vec{p}_i|}
  \begin{bmatrix}
    p_{x,i}^2      & p_{x,i}p_{y,i} & p_{x,i}p_{z,i} \\
    p_{y,i}p_{x,i} & p_{y,i}^2      & p_{y,i}p_{z,i} \\
    p_{z,i}p_{x,i} & p_{z,i}p_{y,i} & p_{z,i}^2
  \end{bmatrix},
  \label{equ:sphericity-tensor}
\end{equation}
where $\vec{p}_i = (p_{x,i}, p_{y,i}, p_{z,i})$ is the three-momentum of particle/jet $i$, and the sum runs over all particles/jets in the region of investigation.
The eigenvalues of the sphericity tensor $\mathcal{M}_{xyz}$ are denoted as $\lambda_1$, $\lambda_2$, and $\lambda_3$ in descending order, i.e., $\lambda_1 \geqslant \lambda_2 \geqslant \lambda_3$ and $\lambda_1 + \lambda_2 + \lambda_3 = 1$.
The transverse sphericity \Sperp\ is then defined as
\begin{equation}
  \Sperp = \frac{2\lambda_2}{\lambda_1 + \lambda_2}.
\end{equation}
In this work, we focus on the configuration of jet-constituent momenta in the plane transverse to the jet axis. However, the tensor defined in Eq.~\eqref{equ:sphericity-tensor} also includes the longitudinal momentum component $p_{z,i}$. 
To characterize only the transverse structure, we project the momenta onto the transverse plane, which is equivalent to setting $p_{z,i} = 0$ when constructing the tensor. 
The resulting transverse sphericity tensor is then represented by a $2 \times 2$ matrix, denoted by $\mathcal{M}_{xy}$,
\begin{equation}
  \mathcal{M}_{xy} = \frac{1}{\sum_i |\vec{p}_{t,i}|}\sum_i \frac{1}{|\vec{p}_{t,i}|}
  \begin{bmatrix}
    p_{x,i}^2      & p_{x,i}p_{y,i} \\
    p_{y,i}p_{x,i} & p_{y,i}^2
  \end{bmatrix}.
  \label{equ:trans-sphericity-tensor}
\end{equation}
The eigenvalues $\lambda_1$ and $\lambda_2$ of $\mathcal{M}_{xy}$, ordered such that $\lambda_1 \geqslant \lambda_2$, satisfy $\lambda_1 + \lambda_2 = 1$.
As illustrated in the right panel of Fig.~\ref{fig:event-shape-illustration}, the transverse sphericity tensor can be represented by a momentum ellipse whose minor semi-axis length is proportional to $\sqrt{\lambda_2}$. 
Consequently, \Sperp\ characterizes the roundness of this ellipse and provides a measure of azimuthal isotropy as encoded by the tensor.
The transverse sphericity \Sperp\ ranges from 0 to 1, where $\Sperp \to 0$ corresponds to a highly collimated, \textit{pencil-like} configuration, while $\Sperp \to 1$ corresponds to a circularly symmetric momentum distribution, as expected for an approximately \textit{isotropic} configuration.

\subsection{Jet frame}
\label{subsec:jet-frame}

\begin{figure}
  \centering
  \includegraphics[width=1\columnwidth,trim=1cm 0 0 0,clip]{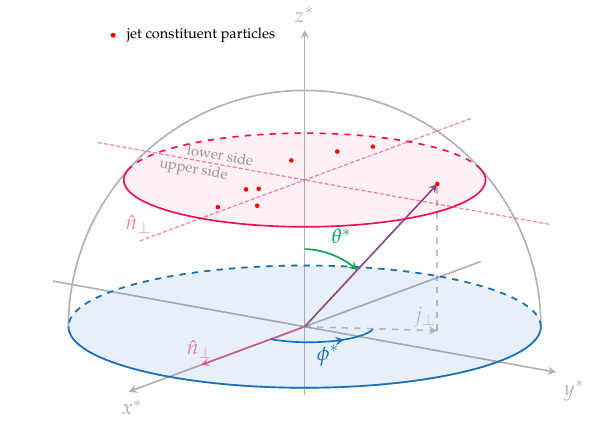}
  \caption{Schematic illustration of jet constituents in the jet frame. The red dots mark the tips of the constituent momentum vectors. For visual clarity, all constituents are drawn with the same longitudinal momentum component $j_z$. The tips of their momentum vectors therefore lie in a common plane perpendicular to the jet axis, represented by the light pink plane. This graphical simplification is not imposed in the analysis.}
\end{figure}

To study the intra-jet shape using event-shape variables, we need to define these observables in the jet frame. The jet frame is a new coordinate frame for each individual jet, where the new $z$-axis is aligned with the jet axis direction. All constituent particles of the jet are then redefined in this new frame, $\vec{p}^{*} = (j_{\perp}, \eta^{*}, \phi^{*})$, where $j_{\perp}$ is the momentum component perpendicular to the jet axis, $\eta^{*} = -\ln[\tan(\theta^{*}/2)]$ is the pseudorapidity in the jet frame, with $\theta^{*}$ being the polar angle relative to the jet axis, and $\phi^{*}$ is the azimuthal angle in the jet frame. The event-shape variables introduced in Sec.~\ref{subsec:evt-shape-def} can then be calculated using the constituent particles' momenta $\vec{p}^{*}$ in the jet frame. The jet frame proposed earlier in Refs.~\cite{Baty:2021ugw,CMS:2023iam} has been used to explore QCD collective evolution within individual jets in \pp~collisions. Furthermore, Ref.~\cite{Yang:2025yaf} investigated the nuclear modification of elliptic anisotropy inside jets calculated in the jet frame, which provides a new perspective for studying jet quenching effects in heavy-ion collisions.
We extend the application of the jet frame to event-shape variables for characterizing the intra-jet shape.
Since the jet direction in the center-of-mass frame varies with respect to the beam axis, directly using the jet constituents' momentum components transverse to the beam axis to calculate intra-jet transverse event-shape variables can introduce a dependence on the jet orientation, thereby introducing geometric effects unrelated to the internal structure of the jet.
By working in the jet frame, we can eliminate the influence of the jet orientation and characterize the geometric and dynamical properties of the parton fragmentation process relative to the initiating parton direction.

To determine the momentum of each constituent particle in the jet frame, we consider a jet with axis direction $\vec{p}$. The new $z$-axis of the jet frame in the center-of-mass (CM) frame is defined as $\hat{z^*} = \hat{\vec{p}} = \vec{p}/|\vec{p}|$. Since the jet axis is generally not aligned with the beam axis (as we typically consider jets within a central rapidity region), we can use the original $z$-axis (beam axis) to determine the new $x$-axis of the jet frame in the CM frame as $\hat{x^*} = \hat{\vec{p}} \times \hat{z}$, and then the new $y$-axis is given by $\hat{y^*} = \hat{\vec{p}} \times \hat{x^*}$. With these three new orthogonal basis vectors, we can easily project the momentum of each constituent particle into the jet frame to obtain $\vec{p}^{*} = (j_{\perp}, \eta^{*}, \phi^{*})$.

\section{Theoretical framework}
\label{sec:framework}

We employ version 2.6.0 of \Jewel~\cite{Zapp:2013vla,Zapp:2012ak,Zapp:2026cqf} to simulate \pp\ collisions and 0--10\% \pbpb\ collisions at \eCMNN{5.02}{TeV}. \Jewel\ is a Monte Carlo event generator designed to simulate QCD jet evolution in the dense medium produced in heavy-ion collisions. It uses a modified version of \textsc{Pythia}~6.4.25~\cite{Sjostrand:2006za} to generate the hard process and initial-state radiation (ISR) and to perform Lund string hadronization. \Jewel\ treats the evolution of final-state parton showers within a common framework in both vacuum and medium simulations. In the medium case, this framework incorporates elastic scatterings between shower partons and thermal medium partons, together with the associated medium-induced radiation.

The virtuality-ordered (or angular-ordered) formalism is well established for generating vacuum parton showers in \pp\ collisions. Unlike approaches that directly modify the splitting functions to include medium effects, \Jewel\ extends this formalism to include medium effects in heavy-ion collisions by applying medium modifications to the parton kinematics between vacuum-like splittings~\cite{Cao:2020wlm}.
The final-state parton shower in \Jewel\ is a virtuality-ordered parton shower formulated in off-shell kinematics. For a parton $a$, the probability for no resolvable splitting to occur between an initial virtuality $Q_{i}^2$ and a lower virtuality $Q_{f}^2$ is given by the Sudakov form factor~\cite{Zapp:2012ak,Zapp:2026cqf}
\begin{widetext}
  \begin{equation}
    \mathcal{S}_a(Q_{i}^2,Q_{f}^2)
    =
    \exp\left[
      -\int_{Q_{f}^2}^{Q_{i}^2}
      \frac{\dd Q'^2}{Q'^2}
      \int_{z_-(Q'^2)}^{z_+(Q'^2)}
      \dd z
      \sum_{b,c}
      \frac{\alpha_s(k_\perp^2)}{2\pi}
      \hat{P}_{a\to bc}(z)
    \right],
  \end{equation}
\end{widetext}
where $\hat{P}_{a\to bc}(z)$ is the unregularized Altarelli--Parisi splitting function for the splitting $a\to b+c$, $z$ is the energy sharing between the partons $b$ and $c$, $\alpha_s(k_\perp^2)$ is the running coupling evaluated at the relative squared transverse momentum $k_\perp^2$ of the splitting. The probability density for a splitting to occur at virtuality $Q^2$ is then given by~\cite{Zapp:2026cqf}
\begin{equation}
  \frac{\dd\mathcal{P}}{\dd Q^2}
  =
  -\frac{\dd\mathcal{S}_a(Q_{i}^2,Q^2)}
  {\dd Q^2}.
\end{equation}
In the \Jewel\ model, the final-state parton shower starts at the hard-process scale and evolves toward lower virtualities through successive emissions until th infrared cutoff is reached~\cite{Zapp:2012ak}.

Propagating shower partons undergo stochastic $2\to 2$ scatterings with thermal scattering centers in the local medium. For an incoming parton $i$ with energy $E$, the local scattering cross section can be written as~\cite{Zapp:2012ak}
\begin{equation}
  \sigma_i^{\mathrm{el}}(E,T)
  =
  \int_0^{|\hat{t}|_{\max}}\!\dd|\hat{t}|
  \int_{x_{\min}}^{x_{\max}}\!\dd x
  \sum_j f_{j}^{\,i}(x,\hat{t})
  \frac{\dd\hat{\sigma}_j}
  {\dd\hat{t}}(x\hat{s},|\hat{t}|).
\end{equation}
Here $f_{j}^{\,i}(x, \hat{t})$ is the effective parton distribution associated with the energetic projectile parton $i$, $|\hat{t}|_{\max}$ is determined by the initial kinematics of the scattering, and $\dd\hat{\sigma}_j/\dd\hat{t}$ is the differential cross section for the $2\to 2$ scattering process. The small-momentum-transfer region is regulated by a Debye screening mass, which is set to $\mu_D\simeq3T$ in this work.
For a parton of energy $E$ that is produced at a branching scale $Q_i^2$ and undergoes a subsequent branching at scale $Q_f^2$, the time interval between the two branchings can be estimated as~\cite{Zapp:2009ud}
\begin{equation}
  \Delta \tau = \frac{E}{Q_f^2} - \frac{E}{Q_i^2}.
\end{equation}
If the parton is produced at time $\tau_i$, the probability for a scattering to occur during this time interval is given by
\begin{equation}
  \mathcal{P}_{\mathrm{scat}}
  =
  1-\exp\left[
    -\int_{\tau_i}^{\tau_i+\Delta\tau}\dd\tau\,
    n(\tau)\,
    \sigma_i^{\mathrm{el}}
  \right],
\end{equation}
where $n(\tau)$ is the local density of scattering centers in the medium.

Medium-induced radiation in \Jewel\ is generated through the parton-shower mechanism described above, rather than being sampled from a separate higher-twist~\cite{Guo:2000nz,Zhang:2003yn,Zhang:2003wk,Zhang:2004qm} gluon-emission spectrum during the subsequent in-medium transport of partons produced by a vacuum shower, as in the LBT~\cite{Wang:2013cia,Li:2010ts,He:2015pra,Cao:2016gvr,Cao:2017hhk} model.
Jet--medium scatterings modify the momentum and virtuality of propagating
partons and can thereby initiate additional shower branchings.
The formation time of a candidate emission is estimated as $\tau_f \sim E/Q^2$.
For a near-collinear gluon emission with energy $\omega$ and transverse momentum $k_\perp$, this estimate is parametrically equivalent to $\tau_f \sim 2\omega/k_\perp^2$.
Momentum transfers from scatterings occurring within the same formation
time cannot be resolved separately by the emission and therefore act coherently.
Their transverse components are added vectorially, and the resulting total transverse momentum transfer modifies $k_\perp$ and hence the formation time.
This procedure is repeated after each additional coherent scattering. Once no further coherent scatterings occur within the updated formation time, the candidate emission is accepted with probability $1/N_{\mathrm{scat}}$, where $N_{\mathrm{scat}}$ is the number of coherent scatterings. This reweighting procedure provides a probabilistic implementation of the non-Abelian Landau--Pomeranchuk--Migdal (LPM) effect~\cite{Zapp:2008af,Zapp:2011ya,Zapp:2012ak}.

For \pbpb\ collisions, we employ the simple medium model distributed with \Jewel.
The QGP is modeled as an ideal gas of quarks and gluons undergoing a longitudinally boost-invariant Bjorken expansion, with a transverse density profile determined from a wounded-nucleon Glauber model~\cite{Zapp:2012ak}.

\section{Results and discussion}
\label{sec:results}

To investigate intra-jet shapes in \pp\ and \pbpb\ collisions, we use \Jewel\ to simulate jet production in \pp\ collisions and in 0--10\% \pbpb\ collisions at \eCMNN{5.02}{TeV}. The \pp\ results provide the vacuum baseline, and a comparison between the two collision systems allows us to study modifications of the intra-jet shape due to jet quenching effects in the QGP medium.

In our analysis, jets are reconstructed from hadrons with $p_t>0.3\,\GeV$ using the anti-$k_t$ algorithm~\cite{Cacciari:2008gp} with the $E$-scheme recombination, implemented in the \FastJet package~\cite{Cacciari:2011ma} with a jet radius parameter $R=0.4$.
We select jets with transverse momentum $\pTjet > 60\GeV$ and pseudorapidity $|\eta^{\mathrm{jet}}| < 1.6$ for further analysis.
For each selected jet, we transform all constituent particle momenta of the jet into the jet frame as described in Sec.~\ref{subsec:jet-frame}, and then compute the event-shape variable defined in Sec.~\ref{subsec:evt-shape-def} using these constituents, thereby characterizing the intra-jet shape.

\begin{figure}[htbp]
  \centering
  \includegraphics[width=0.98\columnwidth]{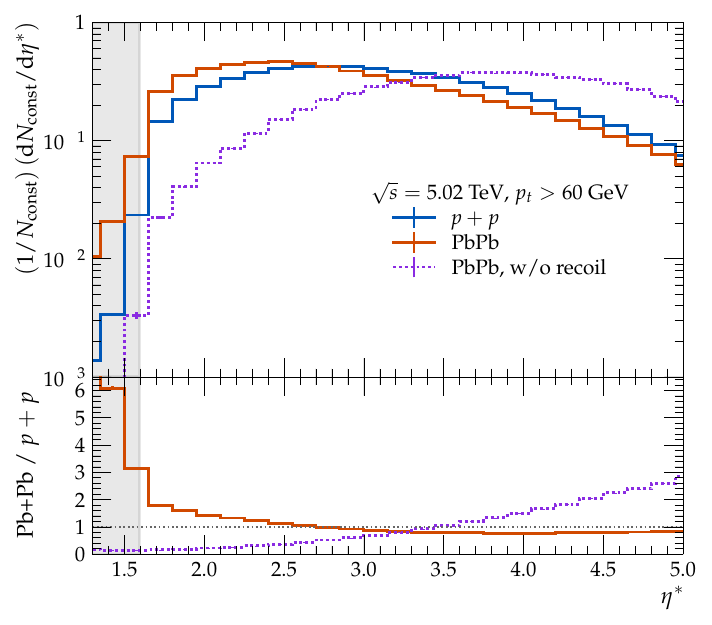}
  \includegraphics[width=0.98\columnwidth]{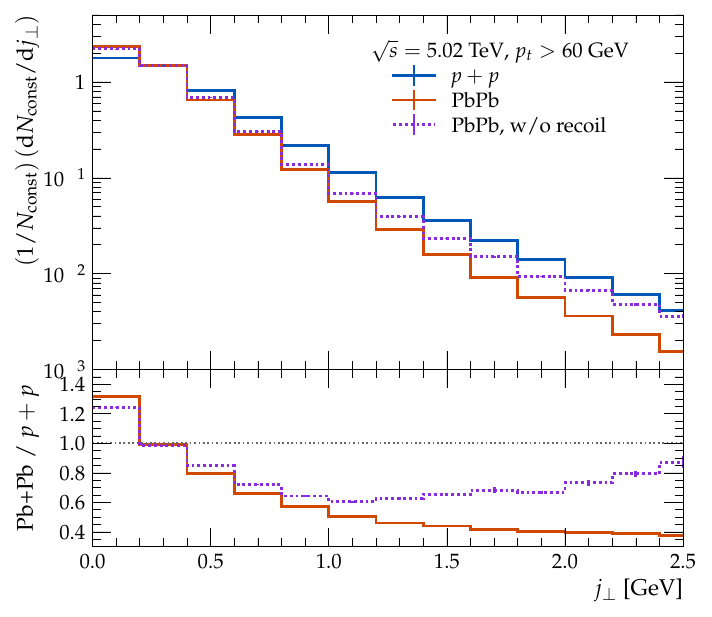}
  \caption{Pseudorapidity $\eta^{*}$ distributions (upper)  and transverse momentum $j_\perp$ distributions (lower) of constituent particles inside jets in the jet frame for \pp\ collisions and 0--10\% central \pbpb\ collisions at 5.02 TeV. The \pbpb-to-\pp\ ratio is shown in the lower panel of each subfigure. The error bars represent the statistical uncertainties. In the $\eta^{*}$ distributions, the shaded bands at low $\eta^{*}$ indicate the region outside the cone with $R=0.4$.}
  \label{fig:etastar-jT-distribution}
\end{figure}

At first, in order to understand the distributions of constituent particles inside jets in the jet-frame in \pp and \pbpb collisions, we present the pseudorapidity $\eta^{*}$ distributions and transverse momentum $j_\perp$ distributions of constituent particles inside jets in the jet frame for both \pp and \pbpb collisions in Fig.~\ref{fig:etastar-jT-distribution}. It should be noted that the $\eta^{*}$ distribution characterizes the angular distance of constituent particles from the jet axis in the jet frame. Here, $\eta^{*} = 0$ corresponds to particles perpendicular to the jet axis, while $\eta^{*} \to \infty$ corresponds to particles collinear with the jet axis. In practice, constituents with $\eta^{*} > 5$ are already nearly collinear with the jet axis.
The shaded bands for $\eta^{*} \leqslant -\ln[\tan(R/2)] \approx 1.60$ indicate the region outside the cone with radius $R=0.4$ in the jet frame.
Jet constituents with angular distances from the jet axis larger than radius $R$ are allowed by the anti-$k_t$ clustering procedure.

\begin{figure*}
  \centering
  \includegraphics[height=0.43\textwidth]{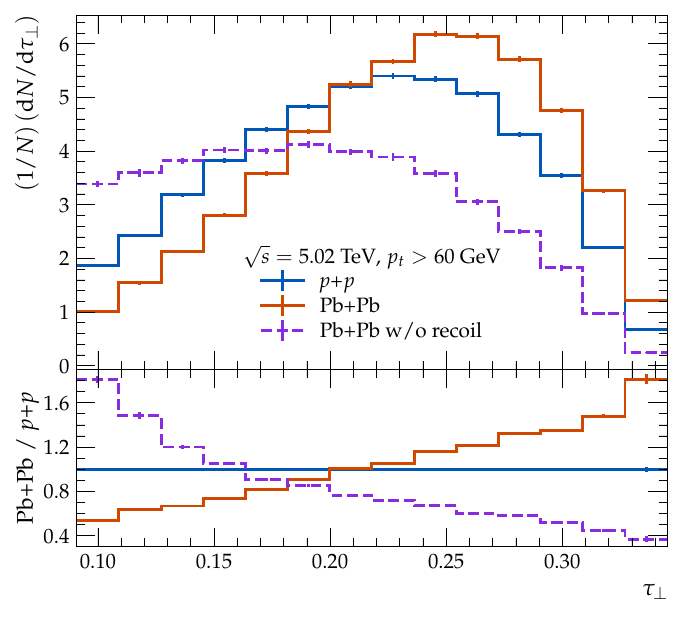}\hfill
  \includegraphics[height=0.43\textwidth]{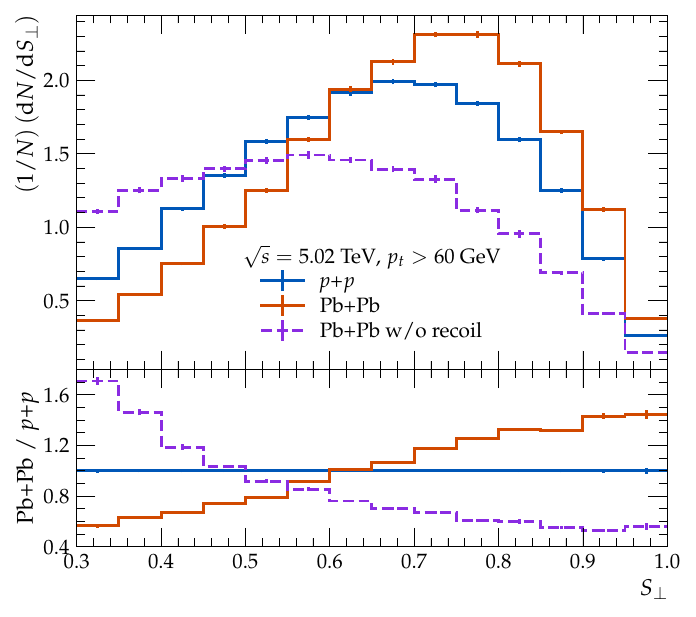}
  \caption{
    Normalized distributions of the intra-jet event-shape observables transverse thrust \TransThrust (left) and transverse sphericity \Sperp (right) for jets with $\pTjet > 60\GeV$ in \pp\ and \pbpb\ collisions at $\sqrt{s_{\mathrm{NN}}}=5.02$ TeV, calculated with \Jewel.
    The blue and orange solid curves show the \pp\ and \pbpb\ results, respectively, with recoil included in the latter.
    The purple dashed curve shows the \pbpb\ result from a calculation without recoil.
    The lower panels show the corresponding \pbpb-to-\pp\ ratios.
  }
  \label{fig:comparison-pp-pbpb-evtshape}
\end{figure*}

As shown in the upper subfigure of Fig.~\ref{fig:etastar-jT-distribution}, the $\eta^{*}$ distributions for jets with $\pTjet > 60\GeV$ in both \pp and \pbpb collisions exhibit similar shapes.
As $\eta^{*}$ increases, corresponding to moving toward the jet axis, the distribution of $\eta^*$ first rises and then falls.
The \pbpb-to-\pp\ ratio shows a relative enhancement of jet constituents at larger angular distances from the jet axis, particularly near the jet-cone boundary, together with a suppression close to the jet axis. This modification is consistent with the picture that jet--medium interactions redistribute jet constituents toward larger angles relative to the jet axis.
In the lower subfigure of Fig.~\ref{fig:etastar-jT-distribution}, the $j_\perp$ distributions in both \pp and \pbpb collisions are shown. The $j_\perp$ distributions exhibit a peak at low $j_\perp$ values, indicating that most jet constituents have small transverse momentum relative to the jet axis. The \pbpb-to-\pp\ ratio shows a slight enhancement at low $j_\perp$ and a suppression at higher $j_\perp$, suggesting that jet--medium interactions lead to a softening of the transverse momentum distribution of jet constituents.

In \Jewel\ simulations of \pbpb\ collisions, a $2\to 2$ scattering between a jet parton and a thermal medium parton can be represented schematically as
\begin{equation}
  p_{\mathrm{jet}}^{\mu}+p_{\mathrm{thermal}}^{\mu}
  \to
  p_{\mathrm{jet}}^{\prime\mu}+p_{\mathrm{recoil}}^{\mu}.
  \label{equ:jewel-scattering}
\end{equation}
In the recoil treatment used in this work, the recoiling medium partons do not undergo further interactions with the medium after the scattering; instead, they free-stream until hadronization. At the hadronization stage, they are converted to gluons, treated as color neighbors of the jet partons with which they interacted, and inserted into the color strings connecting the partons that form the jet.
The resulting hadronic final state therefore includes both the scattered medium partons and the effects of their joint hadronization with the jet shower. This free-streaming treatment represents a limiting case opposite to the assumption of immediate thermalization of recoil energy and momentum made by hydrodynamic frameworks~\cite{KunnawalkamElayavalli:2017hxo}.
In the calculation without recoil, the outgoing medium partons do not participate in hadronization and are not retained in the event record, whereas jet--medium scatterings and medium-associated radiation remain active. As shown in Eq.~\eqref{equ:jewel-scattering}, the recoil four-momentum contains not only the four-momentum $q^{\mu}$ transferred from the jet but also the initial thermal four-momentum $p_{\mathrm{thermal}}^{\mu}$.
The latter originates from the uncorrelated thermal background and must therefore be subtracted before jet observables are evaluated. In this work, we employ the event-level Constituent Subtraction procedure~\cite{Milhano:2022kzx} implemented in \Jewel.

\begin{figure*}
  \centering
  \includegraphics[height=0.43\textwidth]{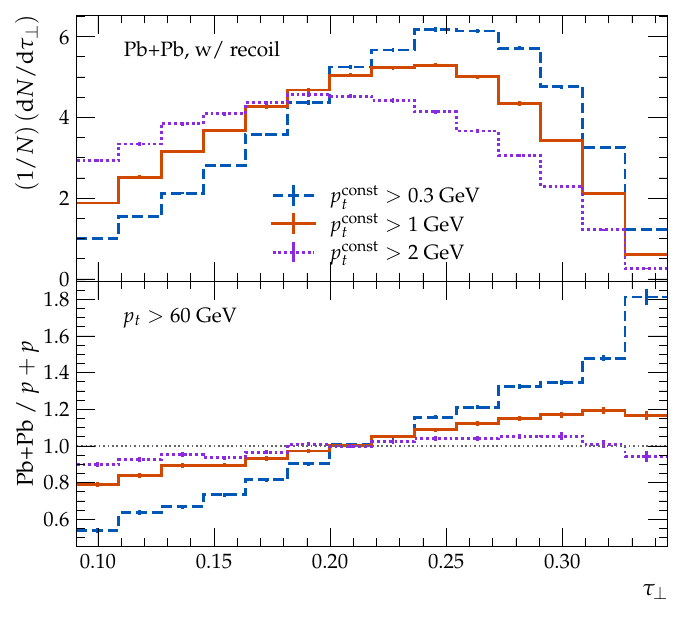}\hfill
  \includegraphics[height=0.43\textwidth]{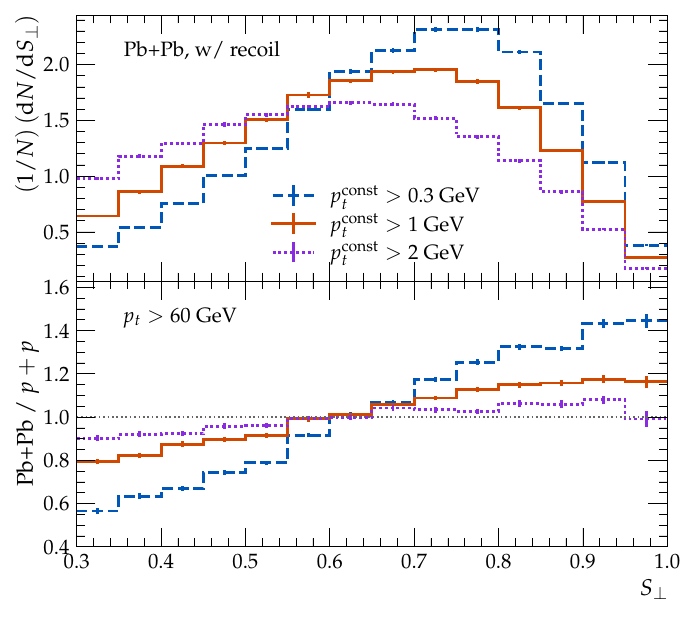}
  \caption{Normalized distributions and corresponding \pbpb-to-\pp\ ratios for the intra-jet shape observables transverse thrust \TransThrust (left) and transverse sphericity \Sperp (right) in the jet frame. The upper panels show the \pbpb distributions for different constituent \pT thresholds, and the lower panels show the corresponding \pbpb-to-\pp\ ratios. Recoil is included in the \pbpb\ calculations. 
  }
  \label{fig:comparison-pp-pbpb-evtshape-ptcut}
\end{figure*}

Figure~\ref{fig:etastar-jT-distribution} also shows a purple dashed curve labeled ``PbPb, w/o recoil'', which represents the \Jewel\ result from a calculation without recoil, in which the recoiling medium partons are not retained in the final state.
We find qualitatively similar modifications of the normalized $j_\perp$ distribution in \pbpb\ collisions relative to \pp\ collisions for the calculations with and without recoil, with both cases exhibiting an enhancement at low $j_\perp$. Therefore, the modifications of the $j_\perp$ distributions do not provide evidence for intra-jet transverse momentum broadening. The $\eta^*$ distribution is modified differently in the calculations with and without recoil. In sharp contrast to the result with recoil, the result without recoil is suppressed at low $\eta^*$ (away from the jet axis) and enhanced at high $\eta^*$ (closer to the jet axis). This means that medium response makes an important contribution to the relative enhancement of soft constituents at the jet periphery.
The results shown in Fig.~\ref{fig:etastar-jT-distribution} are consistent with previous studies showing that jet quenching is accompanied by an increase in the multiplicity of constituents at the periphery of the jet~\cite{KunnawalkamElayavalli:2017hxo}. 
Moreover, the difference between the calculations with and without recoil suggests that medium response makes an important contribution to this relative enhancement at larger angular distances from the jet axis.
This interpretation is consistent with previous studies~\cite{KunnawalkamElayavalli:2017hxo,Chen:2020tbl}.

Figure~\ref{fig:comparison-pp-pbpb-evtshape} shows the intra-jet shape distributions evaluated in the jet frame for \pp\ and \pbpb\ collisions: transverse thrust \TransThrust on the left and transverse sphericity \Sperp on the right. The lower panel of each plot shows the corresponding \pbpb-to-\pp\ ratios.
Compared with \pp, the \pbpb\ calculation with recoil shows a relative enhancement at larger values of both \TransThrust and \Sperp. This trend indicates a less pencil-like and more isotropic distribution of constituent momenta in the plane transverse to the jet axis. 

The modification shown in the left panel of Fig.~\ref{fig:comparison-pp-pbpb-evtshape} suggests that, after jet quenching, the relative weight of constituent transverse momenta away from the transverse thrust axis \TransThrustAxis is increased, making the momentum distribution in the transverse plane of the jet frame less aligned with that axis. This can reduce the maximized momentum projection fraction defined in Eq.~\eqref{equ:Tperp-def} and hence increase \TransThrust.
Similarly, a more isotropic distribution of constituent transverse momenta in the jet frame brings the two normalized eigenvalues of the transverse momentum tensor defined in Eq.~\eqref{equ:trans-sphericity-tensor} closer together, increasing \Sperp. 
The common trend in these two observables characterizes how constituent transverse momenta are distributed among different directions. Since both observables are invariant under a common rescaling of all constituent transverse momenta, their modification depends on changes in the relative momentum weights and angular distribution. They therefore provide information complementary to the single-particle angular and transverse-momentum distributions by characterizing the overall transverse momentum configuration within individual jets. 

Figure~\ref{fig:comparison-pp-pbpb-evtshape} also shows the \pbpb\ calculation without recoil, which exhibits a different modification pattern. In this case, the \pbpb-to-\pp\ ratios show a relative suppression at larger values of both \TransThrust and \Sperp, indicating a more pencil-like and less isotropic distribution of constituent transverse momenta in the jet frame for the selected jets. Thus, the trend toward more isotropic transverse momentum configurations observed in the calculation with recoil is reversed when the recoil contribution is omitted.
The contrast between the two \pbpb\ calculations suggests that the medium response modeled through recoiling medium partons makes an important contribution to the modification of the overall transverse momentum configuration within jets. And that the more isotropic distribution of constituent transverse momenta in the jet frame is mainly due to the contribution of the medium response.

Medium response contributes to the enhancement of soft-particle yields at large angles relative to the jet axis~\cite{KunnawalkamElayavalli:2017hxo}. To further test whether medium response is primarily responsible for the more isotropic distribution of constituent transverse momenta in the jet frame, we apply different \pT cuts to the jet constituents and recalculate the intra-jet shape distributions. The results are shown in Fig.~\ref{fig:comparison-pp-pbpb-evtshape-ptcut}. 
We find that, as the \pT cut on constituent particles increases, the relative enhancement in \pbpb\ collisions compared with \pp\ collisions at larger values of both \TransThrust and \Sperp in \pbpb\ collisions gradually decreases. 
This indicates that the more isotropic distribution of constituent transverse momenta in the jet frame is mainly due to the contribution of soft particles, which are more likely to originate from medium response. 
As the \pT cut increases, the contribution of medium response is gradually suppressed, leading to a less pronounced modification of the intra-jet shape distributions. 
However, increasing the constituent \pT threshold also removes soft particles originating from the jet shower. Raising this threshold is therefore not equivalent to omitting recoil and need not reproduce the relative suppression at larger observable values seen in the calculation without recoil in Fig.~\ref{fig:comparison-pp-pbpb-evtshape}.

\section{Summary}
\label{sec:summary}

In this work, we propose using event-shape variables in the jet frame to characterize the intra-jet momentum configuration in the plane transverse to the jet axis. We then use the \Jewel model to simulate \pp and 0--10\% central \pbpb collisions at \eCMNN{5.02}{TeV}. We study the jet-frame constituent momentum distributions and the distributions of two event-shape observables, transverse thrust \TransThrust and transverse sphericity \Sperp, in both collision systems simulated with \Jewel and investigate their nuclear modifications.

To explain the jet frame, we detail the distributions of pseudorapidity $\eta^{*}$ and transverse momentum $j_\perp$ of constituent particles inside jets in the jet frame for both \pp and \pbpb collisions. We find that medium response contributes to a relative enhancement of soft constituents at the jet periphery in \pbpb collisions.

We further investigate the nuclear modifications of the jet-frame event-shape observables and find that both transverse thrust and transverse sphericity indicate a more isotropic distribution of constituent transverse momenta in the jet frame after jet quenching. Comparing \pbpb calculations with and without recoil, we find that medium response plays an important role in this modification, indicating that both observables are sensitive to medium response effects in jet quenching.

We conclude that the jet-frame event-shape variables provide a novel and effective approach to characterizing the intra-jet shape and studying jet quenching effects in heavy-ion collisions. Other event-shape variables defined in the jet frame can also be explored in future studies to gain further insights into the intra-jet shape and its modifications in the QGP medium.

\section*{Acknowledgments}

We thank W. Dai, Wei-Xi Kong, S.-L. Zhang and W.-J. Xing for valuable discussions and suggestions.
We are grateful to the \textsc{Rivet}~\cite{Bierlich:2024vqo} toolkit;
our measurement code is adapted from its implementation of event-shape variables for hadron/multijet final states.
This work is supported by the National Natural Science Foundation of China (NSFC) under Grant Nos. 12535010 and 12605225.

\section*{Data availability}

This work uses simulated data generated with publicly available software packages. The
simulations and analysis procedures are fully described in this article.

\bibliographystyle{apsrev4-2}
\bibliography{references}

@article{Busza:2018rrf,
    author = "Busza, Wit and Rajagopal, Krishna and van der Schee, Wilke",
    title = "{Heavy Ion Collisions: The Big Picture, and the Big Questions}",
    eprint = "1802.04801",
    archivePrefix = "arXiv",
    primaryClass = "hep-ph",
    reportNumber = "MIT-CTP-4892, MIT-CTP/4892",
    doi = "10.1146/annurev-nucl-101917-020852",
    journal = "Ann. Rev. Nucl. Part. Sci.",
    volume = "68",
    pages = "339--376",
    year = "2018"
}

@article{Zhang:2021xib,
    author = "Zhang, He-Xia and Xiao, Yu-Xin and Kang, Jin-Wen and Zhang, Ben-Wei",
    title = "{Phenomenological study of the anisotropic quark matter in the two-flavor Nambu{\textendash}Jona{\textendash}Lasinio model}",
    eprint = "2102.11792",
    archivePrefix = "arXiv",
    primaryClass = "hep-ph",
    doi = "10.1007/s41365-022-01129-4",
    journal = "Nucl. Sci. Tech.",
    volume = "33",
    number = "11",
    pages = "150",
    year = "2022"
}

@article{Connors:2017ptx,
    author = "Connors, Megan and Nattrass, Christine and Reed, Rosi and Salur, Sevil",
    title = "{Jet measurements in heavy ion physics}",
    eprint = "1705.01974",
    archivePrefix = "arXiv",
    primaryClass = "nucl-ex",
    doi = "10.1103/RevModPhys.90.025005",
    journal = "Rev. Mod. Phys.",
    volume = "90",
    pages = "025005",
    year = "2018"
}

@article{ATLAS:2018gwx,
    author = "Aaboud, Morad and others",
    collaboration = "ATLAS",
    title = "{Measurement of the nuclear modification factor for inclusive jets in Pb+Pb collisions at $\sqrt{s_\mathrm{NN}}=5.02$ TeV with the ATLAS detector}",
    eprint = "1805.05635",
    archivePrefix = "arXiv",
    primaryClass = "nucl-ex",
    reportNumber = "CERN-EP-2018-105",
    doi = "10.1016/j.physletb.2018.10.076",
    journal = "Phys. Lett. B",
    volume = "790",
    pages = "108--128",
    year = "2019"
}

@book{Marzani:2019hun,
    author = "Marzani, Simone and Soyez, Gregory and Spannowsky, Michael",
    title = "Looking inside jets: an introduction to jet substructure and boosted-object phenomenology",
    eprint = "1901.10342",
    archivePrefix = "arXiv",
    primaryClass = "hep-ph",
    doi = "10.1007/978-3-030-15709-8",
    publisher = "Springer",
    address = "Cham",
    volume = "958",
    year = "2019"
}

@article{ALEPH:2003obs,
    author = "Heister, A. and others",
    collaboration = "ALEPH",
    title = "Studies of {QCD} at $e^+ e^-$ centre-of-mass energies between 91~{GeV} and 209~{GeV}",
    reportNumber = "CERN-EP-2003-084",
    doi = "10.1140/epjc/s2004-01891-4",
    journal = "Eur. Phys. J. C",
    volume = "35",
    pages = "457--486",
    year = "2004"
}

@article{CDF:1991etg,
    author = "Abe, F. and others",
    collaboration = "CDF",
    title = "Measurement of {QCD} Jet Broadening in $p\bar{p}$ Collisions at $\sqrt{s} = 1.8$~{TeV}",
    reportNumber = "FERMILAB-PUB-91-023-E, CDF-PUB-JET-PUBLIC-1209, ANL-HEP-PR-91-73",
    doi = "10.1103/PhysRevD.44.601",
    journal = "Phys. Rev. D",
    volume = "44",
    pages = "601--616",
    year = "1991"
}

@article{Vitev:2008rz,
    author = "Vitev, Ivan and Wicks, Simon and Zhang, Ben-Wei",
    title = "{A Theory of jet shapes and cross sections: From hadrons to nuclei}",
    eprint = "0810.2807",
    archivePrefix = "arXiv",
    primaryClass = "hep-ph",
    doi = "10.1088/1126-6708/2008/11/093",
    journal = "JHEP",
    volume = "11",
    pages = "093",
    year = "2008"
}

@article{CDF:1992cus,
    author = "Abe, F. and others",
    collaboration = "CDF",
    title = "{A Measurement of jet shapes in $p\bar{p}$ collisions at $\sqrt{s} = 1.8$ TeV}",
    reportNumber = "FERMILAB-PUB-92-167-E, ANL-HEP-PR-93-46",
    doi = "10.1103/PhysRevLett.70.713",
    journal = "Phys. Rev. Lett.",
    volume = "70",
    pages = "713--717",
    year = "1993"
}

@article{Luo:2018pto,
    author = "Luo, Tan and Cao, Shanshan and He, Yayun and Wang, Xin-Nian",
    title = "{Multiple jets and $\gamma$-jet correlation in high-energy heavy-ion collisions}",
    eprint = "1803.06785",
    archivePrefix = "arXiv",
    primaryClass = "hep-ph",
    doi = "10.1016/j.physletb.2018.06.025",
    journal = "Phys. Lett. B",
    volume = "782",
    pages = "707--716",
    year = "2018"
}

@article{CMS:2018jco,
    author = "Sirunyan, Albert M and others",
    collaboration = "CMS",
    title = "Jet Shapes of Isolated Photon-Tagged Jets in Pb-Pb and $pp$ Collisions at $\sqrt{s_\mathrm{NN}} = 5.02$ {TeV}",
    eprint = "1809.08602",
    archivePrefix = "arXiv",
    primaryClass = "hep-ex",
    reportNumber = "CMS-HIN-18-006, CERN-EP-2018-249",
    doi = "10.1103/PhysRevLett.122.152001",
    journal = "Phys. Rev. Lett.",
    volume = "122",
    number = "15",
    pages = "152001",
    year = "2019"
}

@article{Kang:2023ycg,
    author = "Kang, Jin-Wen and Wang, Sa and Wang, Lei and Zhang, Ben-Wei",
    title = "{Phenomenological study of the angle between jet axes in heavy-ion collisions}",
    eprint = "2312.15518",
    archivePrefix = "arXiv",
    primaryClass = "hep-ph",
    doi = "10.1103/PhysRevC.111.054905",
    journal = "Phys. Rev. C",
    volume = "111",
    number = "5",
    pages = "054905",
    year = "2025"
}

@article{Wang:2024plm,
    author = "Wang, Sa and Li, Yao and Kang, Jin-Wen and Zhang, Ben-Wei",
    title = "{Unveiling the jet angular broadening with photon-tagged jets in high-energy nuclear collisions}",
    eprint = "2408.10924",
    archivePrefix = "arXiv",
    primaryClass = "hep-ph",
    doi = "10.1007/s41365-026-01989-0",
    journal = "Nucl. Sci. Tech.",
    volume = "37",
    number = "9",
    pages = "168",
    year = "2026"
}

@article{Wang:2021jgm,
    author = "Wang, Sa and Kang, Jin-Wen and Dai, Wei and Zhang, Ben-Wei and Wang, Enke",
    title = "Probing the in-medium $P_T$-broadening by $\gamma+\mathrm{HF}$ angular de-correlations",
    eprint = "2107.12000",
    archivePrefix = "arXiv",
    primaryClass = "nucl-th",
    doi = "10.1140/epja/s10050-022-00785-9",
    journal = "Eur. Phys. J. A",
    volume = "58",
    number = "7",
    month = "7",
    pages = "135",
    year = "2022",
    note = "[Erratum: Eur. Phys. J. A 58, 149 (2022)]"
}

@article{He:2011pd,
    author = "He, Yuncun and Vitev, Ivan and Zhang, Ben-Wei",
    title = "${\cal O}(\alpha_s^3)$ Analysis of Inclusive Jet and di-Jet Production in Heavy Ion Reactions at the Large Hadron Collider",
    eprint = "1105.2566",
    archivePrefix = "arXiv",
    primaryClass = "hep-ph",
    doi = "10.1016/j.physletb.2012.05.054",
    journal = "Phys. Lett. B",
    volume = "713",
    pages = "224--232",
    year = "2012"
}

@article{Li:2024pfi,
    author = "Li, Yao and Chen, Shi-Yong and Kong, Wei-Xi and Wang, Sa and Zhang, Ben-Wei",
    title = "{Medium Modifications of Heavy-Flavor Jet Angularities in High-Energy Nuclear Collisions}",
    eprint = "2409.12742",
    archivePrefix = "arXiv",
    primaryClass = "hep-ph",
    doi = "10.1088/0256-307X/42/1/011201",
    journal = "Chin. Phys. Lett.",
    volume = "42",
    number = "1",
    pages = "011201",
    year = "2025"
}

@article{Li:2024uzk,
    author = "Li, Yao and Shen, Shuwan and Wang, Sa and Zhang, Ben-Wei",
    title = "{Transverse momentum balance of dijets in Xe+Xe collisions at the LHC}",
    eprint = "2401.01706",
    archivePrefix = "arXiv",
    primaryClass = "hep-ph",
    doi = "10.1007/s41365-024-01482-6",
    journal = "Nucl. Sci. Tech.",
    volume = "35",
    number = "7",
    pages = "113",
    year = "2024"
}

@article{Dai:2012am,
    author = "Dai, Wei and Vitev, Ivan and Zhang, Ben-Wei",
    title = "{Momentum imbalance of isolated photon-tagged jet production at RHIC and LHC}",
    eprint = "1207.5177",
    archivePrefix = "arXiv",
    primaryClass = "hep-ph",
    doi = "10.1103/PhysRevLett.110.142001",
    journal = "Phys. Rev. Lett.",
    volume = "110",
    number = "14",
    pages = "142001",
    year = "2013"
}

@article{Zhang:2018urd,
    author = "Zhang, Shan-Liang and Luo, Tan and Wang, Xin-Nian and Zhang, Ben-Wei",
    title = "{Z+jet correlation with NLO-matched parton-shower and jet-medium interaction in high-energy nuclear collisions}",
    eprint = "1804.11041",
    archivePrefix = "arXiv",
    primaryClass = "nucl-th",
    doi = "10.1103/PhysRevC.98.021901",
    journal = "Phys. Rev. C",
    volume = "98",
    pages = "021901",
    year = "2018"
}

@article{Caucal:2021cfb,
    author = "Caucal, Paul and Soto-Ontoso, Alba and Takacs, Adam",
    title = "{Dynamically groomed jet radius in heavy-ion collisions}",
    eprint = "2111.14768",
    archivePrefix = "arXiv",
    primaryClass = "hep-ph",
    doi = "10.1103/PhysRevD.105.114046",
    journal = "Phys. Rev. D",
    volume = "105",
    number = "11",
    pages = "114046",
    year = "2022"
}

@article{ALICE:2019ykw,
    author = "Acharya, Shreyasi and others",
    collaboration = "ALICE",
    title = "{Exploration of jet substructure using iterative declustering in pp and Pb{\textendash}Pb collisions at LHC energies}",
    eprint = "1905.02512",
    archivePrefix = "arXiv",
    primaryClass = "nucl-ex",
    reportNumber = "CERN-EP-2019-087",
    doi = "10.1016/j.physletb.2020.135227",
    journal = "Phys. Lett. B",
    volume = "802",
    pages = "135227",
    year = "2020"
}

@article{CMS:2018fof,
    author = "Sirunyan, Albert M and others",
    collaboration = "CMS",
    title = "{Measurement of the groomed jet mass in PbPb and pp collisions at $ \sqrt{s_{\mathrm{NN}}}=5.02 $ TeV}",
    eprint = "1805.05145",
    archivePrefix = "arXiv",
    primaryClass = "hep-ex",
    reportNumber = "CMS-HIN-16-024, CERN-EP-2018-097",
    doi = "10.1007/JHEP10(2018)161",
    journal = "JHEP",
    volume = "10",
    pages = "161",
    year = "2018"
}

@article{STAR:2023pal,
    author = "Aboona, B. E. and others",
    collaboration = "STAR",
    title = "{Measurement of In-Medium Jet Modification Using Direct Photon+Jet and $\pi^0+\mathrm{Jet}$ Correlations in p+p and Central Au+Au Collisions at $\sqrt{s_{\mathrm{NN}}}=200~\mathrm{GeV}$}",
    eprint = "2309.00156",
    archivePrefix = "arXiv",
    primaryClass = "nucl-ex",
    doi = "10.1103/PhysRevLett.134.232301",
    journal = "Phys. Rev. Lett.",
    volume = "134",
    number = "23",
    pages = "232301",
    year = "2025"
}

@article{ATLAS:2018dgb,
    author = "Aaboud, Morad and others",
    collaboration = "ATLAS",
    title = "{Measurement of photon-jet transverse momentum correlations in 5.02 TeV Pb + Pb and $pp$ collisions with ATLAS}",
    eprint = "1809.07280",
    archivePrefix = "arXiv",
    primaryClass = "nucl-ex",
    reportNumber = "CERN-EP-2018-196",
    doi = "10.1016/j.physletb.2018.12.023",
    journal = "Phys. Lett. B",
    volume = "789",
    pages = "167--190",
    year = "2019"
}

@article{Wang:2019xey,
    author = "Wang, Sa and Dai, Wei and Zhang, Ben-Wei and Wang, Enke",
    title = "{Diffusion of charm quarks in jets in high-energy heavy-ion collisions}",
    eprint = "1906.01499",
    archivePrefix = "arXiv",
    primaryClass = "nucl-th",
    doi = "10.1140/epjc/s10052-019-7312-4",
    journal = "Eur. Phys. J. C",
    volume = "79",
    number = "9",
    pages = "789",
    year = "2019"
}

@article{Dasgupta:2003iq,
    author = "Dasgupta, Mrinal and Salam, Gavin P.",
    title = "{Event shapes in e+ e- annihilation and deep inelastic scattering}",
    eprint = "hep-ph/0312283",
    archivePrefix = "arXiv",
    reportNumber = "CERN-TH-2003-306, LPTHE-03-40",
    doi = "10.1088/0954-3899/30/5/R01",
    journal = "J. Phys. G",
    volume = "30",
    pages = "R143",
    year = "2004"
}

@article{ZEUS:2006vwm,
    author = "Chekanov, S. and others",
    collaboration = "ZEUS",
    title = "{Event shapes in deep inelastic scattering at HERA}",
    eprint = "hep-ex/0604032",
    archivePrefix = "arXiv",
    reportNumber = "DESY-06-042",
    doi = "10.1016/j.nuclphysb.2006.05.016",
    journal = "Nucl. Phys. B",
    volume = "767",
    pages = "1--28",
    year = "2007"
}

@article{CMS:2014tkl,
    author = "Khachatryan, Vardan and others",
    collaboration = "CMS",
    title = "{Study of Hadronic Event-Shape Variables in Multijet Final States in pp Collisions at $\sqrt{s}$ = 7 TeV}",
    eprint = "1407.2856",
    archivePrefix = "arXiv",
    primaryClass = "hep-ex",
    reportNumber = "CMS-SMP-12-022, CERN-PH-EP-2014-146",
    doi = "10.1007/JHEP10(2014)087",
    journal = "JHEP",
    volume = "10",
    pages = "087",
    year = "2014"
}

@article{CMS:2025sws,
    author = "Chekhovsky, Vladimir and others",
    collaboration = "CMS",
    title = "{Measurement of event shapes in minimum-bias events from proton-proton collisions at s=13{\,}{\,}TeV}",
    eprint = "2505.17850",
    archivePrefix = "arXiv",
    primaryClass = "hep-ex",
    reportNumber = "CMS-SMP-23-008, CERN-EP-2025-041",
    doi = "10.1103/rqvb-gfb7",
    journal = "Phys. Rev. D",
    volume = "112",
    number = "11",
    pages = "112006",
    year = "2025"
}

@article{ATLAS:2012uka,
    author = "Aad, Georges and others",
    collaboration = "ATLAS",
    title = "{Measurement of charged-particle event shape variables in $\sqrt{s}=7$ TeV proton-proton interactions with the ATLAS detector}",
    eprint = "1207.6915",
    archivePrefix = "arXiv",
    primaryClass = "hep-ex",
    reportNumber = "CERN-PH-EP-2012-178",
    doi = "10.1103/PhysRevD.88.032004",
    journal = "Phys. Rev. D",
    volume = "88",
    number = "3",
    pages = "032004",
    year = "2013"
}

@article{ALICE:2012cor,
    author = "Abelev, Betty and others",
    collaboration = "ALICE",
    title = "{Transverse sphericity of primary charged particles in minimum bias proton-proton collisions at $\sqrt{s}=0.9$, 2.76 and 7 TeV}",
    eprint = "1205.3963",
    archivePrefix = "arXiv",
    primaryClass = "hep-ex",
    reportNumber = "CERN-PH-EP-2012-136",
    doi = "10.1140/epjc/s10052-012-2124-9",
    journal = "Eur. Phys. J. C",
    volume = "72",
    pages = "2124",
    year = "2012"
}

@article{ATLAS:2012tch,
    author = "Aad, Georges and others",
    collaboration = "ATLAS",
    title = "{Measurement of event shapes at large momentum transfer with the ATLAS detector in $pp$ collisions at $\sqrt{s}=7$ TeV}",
    eprint = "1206.2135",
    archivePrefix = "arXiv",
    primaryClass = "hep-ex",
    reportNumber = "CERN-PH-EP-2012-119, SLAC-PUB-16021",
    doi = "10.1140/epjc/s10052-012-2211-y",
    journal = "Eur. Phys. J. C",
    volume = "72",
    pages = "2211",
    year = "2012"
}

@article{CMS:2018svp,
    author = "Sirunyan, Albert M and others",
    collaboration = "CMS",
    title = "{Event shape variables measured using multijet final states in proton-proton collisions at $ \sqrt{s}=13 $ TeV}",
    eprint = "1811.00588",
    archivePrefix = "arXiv",
    primaryClass = "hep-ex",
    reportNumber = "CMS-SMP-17-003, CERN-EP-2018-253",
    doi = "10.1007/JHEP12(2018)117",
    journal = "JHEP",
    volume = "12",
    pages = "117",
    year = "2018"
}

@article{ATLAS:2020vup,
    author = "Aad, Georges and others",
    collaboration = "ATLAS",
    title = "{Measurement of hadronic event shapes in high-p$_{T}$ multijet final states at $ \sqrt{s} $ = 13 TeV with the ATLAS detector}",
    eprint = "2007.12600",
    archivePrefix = "arXiv",
    primaryClass = "hep-ex",
    reportNumber = "CERN-EP-2020-077",
    doi = "10.1007/JHEP01(2021)188",
    journal = "JHEP",
    volume = "01",
    pages = "188",
    year = "2021",
    note = "[Erratum: JHEP 12, 053 (2021)]"
}

@article{Chen:2020pfa,
    author = "Chen, Shi-Yong and Dai, Wei and Zhang, Shan-Liang and Zhang, Qing and Zhang, Ben-Wei",
    title = "{The global geometrical property of jet events in high-energy nuclear collisions}",
    eprint = "2005.02892",
    archivePrefix = "arXiv",
    primaryClass = "hep-ph",
    doi = "10.1140/epjc/s10052-020-08452-3",
    journal = "Eur. Phys. J. C",
    volume = "80",
    number = "9",
    pages = "865",
    year = "2020"
}

@article{Kang:2023qxb,
    author = "Kang, Jin-Wen and Wang, Lei and Dai, Wei and Wang, Sa and Zhang, Ben-Wei",
    title = "{Multijet topology in high-energy nuclear collisions: Jet broadening}",
    eprint = "2304.04649",
    archivePrefix = "arXiv",
    primaryClass = "nucl-th",
    doi = "10.1103/71sd-7qqb",
    journal = "Phys. Rev. C",
    volume = "112",
    number = "3",
    pages = "034903",
    year = "2025"
}

@article{Kong:2024wdk,
    author = "Kong, Wei-Xi and Zhang, Ben-Wei",
    title = "{Fox-Wolfram moment of jet production in relativistic heavy-ion collisions*}",
    eprint = "2407.20680",
    archivePrefix = "arXiv",
    primaryClass = "hep-ph",
    doi = "10.1088/1674-1137/add5d8",
    journal = "Chin. Phys. C",
    volume = "49",
    number = "9",
    pages = "094101",
    year = "2025"
}

@article{Banfi:2010xy,
    author = "Banfi, Andrea and Salam, Gavin P. and Zanderighi, Giulia",
    title = "{Phenomenology of event shapes at hadron colliders}",
    eprint = "1001.4082",
    archivePrefix = "arXiv",
    primaryClass = "hep-ph",
    doi = "10.1007/JHEP06(2010)038",
    journal = "JHEP",
    volume = "06",
    pages = "038",
    year = "2010"
}

@article{Cesarotti:2020hwb,
    author = "Cesarotti, Cari and Thaler, Jesse",
    title = "{A Robust Measure of Event Isotropy at Colliders}",
    eprint = "2004.06125",
    archivePrefix = "arXiv",
    primaryClass = "hep-ph",
    reportNumber = "MIT-CTP 5195",
    doi = "10.1007/JHEP08(2020)084",
    journal = "JHEP",
    volume = "08",
    pages = "084",
    year = "2020"
}

@article{Cesarotti:2020ngq,
    author = "Cesarotti, Cari and Reece, Matthew and Strassler, Matthew J.",
    title = "{The efficacy of event isotropy as an event shape observable}",
    eprint = "2011.06599",
    archivePrefix = "arXiv",
    primaryClass = "hep-ph",
    doi = "10.1007/JHEP07(2021)215",
    journal = "JHEP",
    volume = "07",
    pages = "215",
    year = "2021"
}

@article{Prasad:2025yfj,
    author = "Prasad, Suraj and Tripathy, Sushanta and Sahoo, Bhagyarathi and Sahoo, Raghunath",
    title = "{Event topology classifiers at the Large Hadron Collider}",
    eprint = "2506.03782",
    archivePrefix = "arXiv",
    primaryClass = "hep-ph",
    doi = "10.1016/j.physrep.2026.04.001",
    journal = "Phys. Rept.",
    volume = "1181",
    pages = "1--75",
    year = "2026"
}

@article{Okorokov:2011mb,
    author = "Okorokov, V. A.",
    title = "{Event Shapes for Hadronic Final State: Experimental Review}",
    eprint = "1105.4784",
    archivePrefix = "arXiv",
    primaryClass = "hep-ph",
    doi = "10.1142/S0217751X12500376",
    journal = "Int. J. Mod. Phys. A",
    volume = "27",
    pages = "1250037",
    year = "2012"
}

@article{Mehtar-Tani:2025rty,
    author = "Mehtar-Tani, Yacine",
    title = "{The Physics of Jet Quenching in Perturbative QCD}",
    eprint = "2509.26394",
    archivePrefix = "arXiv",
    primaryClass = "hep-ph",
    journal = ""
}

@article{Wang:1994fx,
    author = "Wang, Xin-Nian and Gyulassy, Miklos and Plumer, Michael",
    title = "{The LPM effect in QCD and radiative energy loss in a quark gluon plasma}",
    eprint = "hep-ph/9408344",
    archivePrefix = "arXiv",
    reportNumber = "LBL-35980",
    doi = "10.1103/PhysRevD.51.3436",
    journal = "Phys. Rev. D",
    volume = "51",
    pages = "3436--3446",
    year = "1995",
    inspirehepID = "375799"
}

@article{Gyulassy:2000er,
    author = "Gyulassy, M. and Levai, P. and Vitev, I.",
    title = "{Reaction operator approach to nonAbelian energy loss}",
    eprint = "nucl-th/0006010",
    archivePrefix = "arXiv",
    reportNumber = "CU-TP-979",
    doi = "10.1016/S0550-3213(00)00652-0",
    journal = "Nucl. Phys. B",
    volume = "594",
    pages = "371--419",
    year = "2001",
    inspirehepID = "528435"
}

@article{Arnold:2001ms,
    author = "Arnold, Peter Brockway and Moore, Guy D. and Yaffe, Laurence G.",
    title = "{Photon emission from quark gluon plasma: Complete leading order results}",
    eprint = "hep-ph/0111107",
    archivePrefix = "arXiv",
    reportNumber = "UW-PT-01-22",
    doi = "10.1088/1126-6708/2001/12/009",
    journal = "JHEP",
    volume = "12",
    pages = "009",
    year = "2001",
    inspirehepID = "566142"
}

@article{Baier:1996kr,
    author = "Baier, R. and Dokshitzer, Yuri L. and Mueller, Alfred H. and Peigne, S. and Schiff, D.",
    title = "{Radiative energy loss of high-energy quarks and gluons in a finite volume quark - gluon plasma}",
    eprint = "hep-ph/9607355",
    archivePrefix = "arXiv",
    reportNumber = "BI-TP-96-21, CUTP-759, LPTHE-ORSAY-96-34",
    doi = "10.1016/S0550-3213(96)00553-6",
    journal = "Nucl. Phys. B",
    volume = "483",
    pages = "291--320",
    year = "1997",
    inspirehepID = "420877"
}

@article{Zakharov:1997uu,
    author = "Zakharov, B. G.",
    title = "{Radiative energy loss of high-energy quarks in finite size nuclear matter and quark - gluon plasma}",
    eprint = "hep-ph/9704255",
    archivePrefix = "arXiv",
    reportNumber = "LPTHE-ORSAY-97-09",
    doi = "10.1134/1.567389",
    journal = "JETP Lett.",
    volume = "65",
    pages = "615--620",
    year = "1997",
    inspirehepID = "441901"
}

@article{Wiedemann:2000za,
    author = "Wiedemann, Urs Achim",
    title = "{Gluon radiation off hard quarks in a nuclear environment: Opacity expansion}",
    eprint = "hep-ph/0005129",
    archivePrefix = "arXiv",
    doi = "10.1016/S0550-3213(00)00457-0",
    journal = "Nucl. Phys. B",
    volume = "588",
    pages = "303--344",
    year = "2000",
    inspirehepID = "527344"
}

@article{Guo:2000nz,
    author = "Guo, Xiao-feng and Wang, Xin-Nian",
    title = "{Multiple scattering, parton energy loss and modified fragmentation functions in deeply inelastic e A scattering}",
    eprint = "hep-ph/0005044",
    archivePrefix = "arXiv",
    reportNumber = "LBNL-45631, LBL-45631",
    doi = "10.1103/PhysRevLett.85.3591",
    journal = "Phys. Rev. Lett.",
    volume = "85",
    pages = "3591--3594",
    year = "2000",
    inspirehepID = "526889"
}

@article{Zhang:2003wk,
    author = "Zhang, Ben-Wei and Wang, Enke and Wang, Xin-Nian",
    title = "{Heavy quark energy loss in nuclear medium}",
    eprint = "nucl-th/0309040",
    archivePrefix = "arXiv",
    reportNumber = "LBNL-53791",
    doi = "10.1103/PhysRevLett.93.072301",
    journal = "Phys. Rev. Lett.",
    volume = "93",
    pages = "072301",
    year = "2004",
    inspirehepID = "628366"
}

@article{Vitev:2009rd,
    author = "Vitev, Ivan and Zhang, Ben-Wei",
    title = "{Jet tomography of high-energy nucleus-nucleus collisions at next-to-leading order}",
    eprint = "0910.1090",
    archivePrefix = "arXiv",
    primaryClass = "hep-ph",
    doi = "10.1103/PhysRevLett.104.132001",
    journal = "Phys. Rev. Lett.",
    volume = "104",
    pages = "132001",
    year = "2010"
}

@article{Qin:2015srf,
    author = "Qin, Guang-You and Wang, Xin-Nian",
    editor = "Wang, Xin-Nian",
    title = "{Jet quenching in high-energy heavy-ion collisions}",
    eprint = "1511.00790",
    archivePrefix = "arXiv",
    primaryClass = "hep-ph",
    doi = "10.1142/S0218301315300143",
    journal = "Int. J. Mod. Phys. E",
    volume = "24",
    number = "11",
    pages = "1530014",
    year = "2015"
}

@article{Cao:2024pxc,
    author = "Cao, Shanshan and Majumder, Abhijit and Modarresi-Yazdi, Rouzbeh and Soudi, Ismail and Tachibana, Yasuki",
    title = "{Jet quenching: From theory to simulation}",
    eprint = "2401.10026",
    archivePrefix = "arXiv",
    primaryClass = "hep-ph",
    doi = "10.1142/9789811294679_0001",
    journal = "Int. J. Mod. Phys. E",
    volume = "33",
    number = "08",
    pages = "2430002",
    year = "2024"
}

@article{Counihan:1975rr,
    author = "Counihan, M. J.",
    title = "{`Principal Axis' Variables for the Analysis of Multiparticle Production}",
    reportNumber = "CERN/D.Ph.II/PHYS 75-43",
    doi = "10.1016/0370-2693(75)90093-3",
    journal = "Phys. Lett. B",
    volume = "59",
    pages = "367--368",
    year = "1975"
}

@article{ATHENS-LIVERPOOL-VIENNA:1977fca,
    author = "Kakoulidou, M. and others",
    collaboration = "ATHENS-LIVERPOOL-VIENNA",
    title = "{Study of the Planarity in K- p Reactions at 8.25-GeV/c}",
    doi = "10.1016/0550-3213(77)90103-1",
    journal = "Nucl. Phys. B",
    volume = "130",
    pages = "189--208",
    year = "1977"
}

@article{Parisi:1978eg,
    author = "Parisi, G.",
    title = "{Super Inclusive Cross-Sections}",
    reportNumber = "LPTENS 78/5",
    doi = "10.1016/0370-2693(78)90061-8",
    journal = "Phys. Lett. B",
    volume = "74",
    pages = "65--67",
    year = "1978"
}

@article{Donoghue:1979vi,
    author = "Donoghue, John F. and Low, F. E. and Pi, So-Young",
    title = "{Tensor Analysis of Hadronic Jets in Quantum Chromodynamics}",
    reportNumber = "MIT-CTP-771",
    doi = "10.1103/PhysRevD.20.2759",
    journal = "Phys. Rev. D",
    volume = "20",
    pages = "2759",
    year = "1979"
}

@article{Baty:2021ugw,
    author = "Baty, Austin and Gardner, Parker and Li, Wei",
    title = "{Novel observables for exploring QCD collective evolution and quantum entanglement within individual jets}",
    eprint = "2104.11735",
    archivePrefix = "arXiv",
    primaryClass = "hep-ph",
    doi = "10.1103/PhysRevC.107.064908",
    journal = "Phys. Rev. C",
    volume = "107",
    number = "6",
    pages = "064908",
    year = "2023"
}

@article{CMS:2023iam,
    author = "Hayrapetyan, Aram and others",
    collaboration = "CMS",
    title = "{Observation of Enhanced Long-Range Elliptic Anisotropies Inside High-Multiplicity Jets in pp Collisions at s=13{\,}{\,}TeV}",
    eprint = "2312.17103",
    archivePrefix = "arXiv",
    primaryClass = "hep-ex",
    reportNumber = "CMS-HIN-21-013, CERN-EP-2023-281",
    doi = "10.1103/PhysRevLett.133.142301",
    journal = "Phys. Rev. Lett.",
    volume = "133",
    number = "14",
    pages = "142301",
    year = "2024"
}

@article{Yang:2025yaf,
    author = "Yang, Meng-Quan and Kong, Wei-Xi and Ru, Peng and Zhang, Ben-Wei",
    title = "{Suppression of elliptic anisotropy inside jets: A new perspective for jet quenching}",
    eprint = "2503.15042",
    archivePrefix = "arXiv",
    primaryClass = "hep-ph",
    doi = "10.1016/j.physletb.2025.139746",
    journal = "Phys. Lett. B",
    volume = "868",
    pages = "139746",
    year = "2025"
}

@article{Wang:2013cia,
    author = "Wang, Xin-Nian and Zhu, Yan",
    title = "{Medium Modification of $\gamma$-jets in High-energy Heavy-ion Collisions}",
    eprint = "1302.5874",
    archivePrefix = "arXiv",
    primaryClass = "hep-ph",
    doi = "10.1103/PhysRevLett.111.062301",
    journal = "Phys. Rev. Lett.",
    volume = "111",
    number = "6",
    pages = "062301",
    year = "2013"
}

@article{Li:2010ts,
    author = "Li, Hanlin and Liu, Fuming and Ma, Guo-liang and Wang, Xin-Nian and Zhu, Yan",
    title = "{Mach cone induced by $\gamma$-triggered jets in high-energy heavy-ion collisions}",
    eprint = "1006.2893",
    archivePrefix = "arXiv",
    primaryClass = "nucl-th",
    doi = "10.1103/PhysRevLett.106.012301",
    journal = "Phys. Rev. Lett.",
    volume = "106",
    pages = "012301",
    year = "2011"
}

@article{He:2015pra,
    author = "He, Yayun and Luo, Tan and Wang, Xin-Nian and Zhu, Yan",
    title = "{Linear Boltzmann Transport for Jet Propagation in the Quark-Gluon Plasma: Elastic Processes and Medium Recoil}",
    eprint = "1503.03313",
    archivePrefix = "arXiv",
    primaryClass = "nucl-th",
    doi = "10.1103/PhysRevC.91.054908",
    journal = "Phys. Rev. C",
    volume = "91",
    pages = "054908",
    year = "2015",
    note = "[Erratum: Phys.Rev.C 97, 019902 (2018)]"
}

@article{Cao:2016gvr,
    author = "Cao, Shanshan and Luo, Tan and Qin, Guang-You and Wang, Xin-Nian",
    title = "{Linearized Boltzmann transport model for jet propagation in the quark-gluon plasma: Heavy quark evolution}",
    eprint = "1605.06447",
    archivePrefix = "arXiv",
    primaryClass = "nucl-th",
    doi = "10.1103/PhysRevC.94.014909",
    journal = "Phys. Rev. C",
    volume = "94",
    number = "1",
    pages = "014909",
    year = "2016"
}

@article{Cao:2017hhk,
    author = "Cao, Shanshan and Luo, Tan and Qin, Guang-You and Wang, Xin-Nian",
    title = "{Heavy and light flavor jet quenching at RHIC and LHC energies}",
    eprint = "1703.00822",
    archivePrefix = "arXiv",
    primaryClass = "nucl-th",
    doi = "10.1016/j.physletb.2017.12.023",
    journal = "Phys. Lett. B",
    volume = "777",
    pages = "255--259",
    year = "2018"
}

@article{Zhang:2003yn,
    author = "Zhang, Ben-Wei and Wang, Xin-Nian",
    title = "{Multiple parton scattering in nuclei: Beyond helicity amplitude approximation}",
    eprint = "hep-ph/0301195",
    archivePrefix = "arXiv",
    reportNumber = "LBNL-52041",
    doi = "10.1016/S0375-9474(03)01003-0",
    journal = "Nucl. Phys. A",
    volume = "720",
    pages = "429--451",
    year = "2003"
}

@article{Zhang:2004qm,
    author = "Zhang, Ben-Wei and Wang, En-ke and Wang, Xin-Nian",
    title = "{Multiple parton scattering in nuclei: Heavy quark energy loss and modified fragmentation functions}",
    eprint = "hep-ph/0412060",
    archivePrefix = "arXiv",
    reportNumber = "LBNL-56660",
    doi = "10.1016/j.nuclphysa.2005.04.022",
    journal = "Nucl. Phys. A",
    volume = "757",
    pages = "493--524",
    year = "2005"
}

@article{Cacciari:2008gp,
    author = "Cacciari, Matteo and Salam, Gavin P. and Soyez, Gregory",
    title = "{The anti-$k_t$ jet clustering algorithm}",
    eprint = "0802.1189",
    archivePrefix = "arXiv",
    primaryClass = "hep-ph",
    reportNumber = "LPTHE-07-03",
    doi = "10.1088/1126-6708/2008/04/063",
    journal = "JHEP",
    volume = "04",
    pages = "063",
    year = "2008"
}

@article{Cacciari:2011ma,
    author = "Cacciari, Matteo and Salam, Gavin P. and Soyez, Gregory",
    title = "{FastJet User Manual}",
    eprint = "1111.6097",
    archivePrefix = "arXiv",
    primaryClass = "hep-ph",
    reportNumber = "CERN-PH-TH-2011-297",
    doi = "10.1140/epjc/s10052-012-1896-2",
    journal = "Eur. Phys. J. C",
    volume = "72",
    pages = "1896",
    year = "2012"
}

@article{Bierlich:2024vqo,
    author = "Bierlich, Christian and Buckley, Andy and Butterworth, Jonathan Mark and Gutschow, Christian and Lonnblad, Leif and Procter, Tomasz and Richardson, Peter and Yeh, Yoran",
    title = "{Robust independent validation of experiment and theory: Rivet version 4 release note}",
    eprint = "2404.15984",
    archivePrefix = "arXiv",
    primaryClass = "hep-ph",
    reportNumber = "MCNET-24-05",
    doi = "10.21468/SciPostPhysCodeb.36",
    journal = "SciPost Phys. Codeb.",
    volume = "36",
    pages = "1",
    year = "2024"
}

@article{Zapp:2013vla,
    author = "Zapp, Korinna C.",
    title = "{JEWEL 2.0.0: directions for use}",
    eprint = "1311.0048",
    archivePrefix = "arXiv",
    primaryClass = "hep-ph",
    reportNumber = "CERN-PH-TH-2013-255, MCNET-13-17",
    doi = "10.1140/epjc/s10052-014-2762-1",
    journal = "Eur. Phys. J. C",
    volume = "74",
    number = "2",
    pages = "2762",
    year = "2014"
}

@article{Zapp:2012ak,
    author = "Zapp, Korinna C. and Krauss, Frank and Wiedemann, Urs A.",
    title = "{A perturbative framework for jet quenching}",
    eprint = "1212.1599",
    archivePrefix = "arXiv",
    primaryClass = "hep-ph",
    reportNumber = "CERN-PH-TH-2012-344, IPPP-12-92, DCPT-12-84, MCNET-12-14",
    doi = "10.1007/JHEP03(2013)080",
    journal = "JHEP",
    volume = "03",
    pages = "080",
    year = "2013"
}

@article{Zapp:2026cqf,
    author = "Zapp, Korinna",
    title = "{A dynamical implementation of colour coherence for quenched jets in JEWEL}",
    eprint = "2604.13932",
    archivePrefix = "arXiv",
    primaryClass = "hep-ph",
    reportNumber = "MCNET-26-06",
    month = "4",
    year = "2026",
    journal = ""
}

@article{Sjostrand:2006za,
    author = "Sjostrand, Torbjorn and Mrenna, Stephen and Skands, Peter Z.",
    title = "{PYTHIA 6.4 Physics and Manual}",
    eprint = "hep-ph/0603175",
    archivePrefix = "arXiv",
    reportNumber = "FERMILAB-PUB-06-052-CD-T, LU-TP-06-13",
    doi = "10.1088/1126-6708/2006/05/026",
    journal = "JHEP",
    volume = "05",
    pages = "026",
    year = "2006"
}

@article{Cao:2020wlm,
    author = "Cao, Shanshan and Wang, Xin-Nian",
    title = "{Jet quenching and medium response in high-energy heavy-ion collisions: a review}",
    eprint = "2002.04028",
    archivePrefix = "arXiv",
    primaryClass = "hep-ph",
    doi = "10.1088/1361-6633/abc22b",
    journal = "Rept. Prog. Phys.",
    volume = "84",
    number = "2",
    pages = "024301",
    year = "2021"
}

@article{Zapp:2009ud,
    author = "Zapp, Korinna and Stachel, Johanna and Wiedemann, Urs Achim",
    editor = "Eskola, Kari J. and L{\'e}vai, Peter and Morsch, Andreas and Petracek, Vojtech and Rak, Jan",
    title = "{JEWEL - a Monte Carlo Model for Jet Quenching}",
    eprint = "0904.4885",
    archivePrefix = "arXiv",
    primaryClass = "hep-ph",
    doi = "10.22323/1.080.0022",
    journal = "PoS",
    volume = "High-pT physics09",
    pages = "022",
    year = "2009"
}

@article{Zapp:2008af,
    author = "Zapp, Korinna and Stachel, Johanna and Wiedemann, Urs Achim",
    title = "{A Local Monte Carlo implementation of the non-abelian Landau-Pomerantschuk-Migdal effect}",
    eprint = "0812.3888",
    archivePrefix = "arXiv",
    primaryClass = "hep-ph",
    reportNumber = "CERN-PH-TH-2008-244",
    doi = "10.1103/PhysRevLett.103.152302",
    journal = "Phys. Rev. Lett.",
    volume = "103",
    pages = "152302",
    year = "2009"
}

@article{Zapp:2011ya,
    author = "Zapp, Korinna Christine and Stachel, Johanna and Wiedemann, Urs Achim",
    title = "{A local Monte Carlo framework for coherent QCD parton energy loss}",
    eprint = "1103.6252",
    archivePrefix = "arXiv",
    primaryClass = "hep-ph",
    reportNumber = "IPPP-11-16, DCPT-11-32, MCNET-11-11, CERN-PH-TH-2011-043",
    doi = "10.1007/JHEP07(2011)118",
    journal = "JHEP",
    volume = "07",
    pages = "118",
    year = "2011"
}

@article{Milhano:2022kzx,
    author = "Milhano, Jos{\'e} Guilherme and Zapp, Korinna",
    title = "{Improved background subtraction and a fresh look at jet sub-structure in JEWEL}",
    eprint = "2207.14814",
    archivePrefix = "arXiv",
    primaryClass = "hep-ph",
    reportNumber = "LU-TH 22-48, MCNET-22-12",
    doi = "10.1140/epjc/s10052-022-10954-1",
    journal = "Eur. Phys. J. C",
    volume = "82",
    number = "11",
    pages = "1010",
    year = "2022"
}

@article{KunnawalkamElayavalli:2017hxo,
    author = "Kunnawalkam Elayavalli, Raghav and Zapp, Korinna Christine",
    title = "{Medium response in JEWEL and its impact on jet shape observables in heavy ion collisions}",
    eprint = "1707.01539",
    archivePrefix = "arXiv",
    primaryClass = "hep-ph",
    reportNumber = "CERN-TH-2017-146, MCNET-17-11",
    doi = "10.1007/JHEP07(2017)141",
    journal = "JHEP",
    volume = "07",
    pages = "141",
    year = "2017"
}

@article{Chen:2020tbl,
    author = "Chen, Wei and Cao, Shanshan and Luo, Tan and Pang, Long-Gang and Wang, Xin-Nian",
    title = "{Medium modification of $\gamma$-jet fragmentation functions in Pb+Pb collisions at LHC}",
    eprint = "2005.09678",
    archivePrefix = "arXiv",
    primaryClass = "hep-ph",
    doi = "10.1016/j.physletb.2020.135783",
    journal = "Phys. Lett. B",
    volume = "810",
    pages = "135783",
    year = "2020"
}

@article{Wang:2025mkw,
    author = "Wang, Sa and Li, Shuang and Kang, Jin-Wen and Zhang, Ben-Wei and Wang, Enke",
    title = "{Substructure grooming of inclusive and photon-tagged jets in heavy-ion collisions}",
    eprint = "2512.14053",
    archivePrefix = "arXiv",
    primaryClass = "hep-ph",
    doi = "10.1007/s11433-026-3057-4",
    journal = "Sci. China Phys. Mech. Astron.",
    volume = "69",
    number = "11",
    pages = "111011",
    year = "2026"
}

\end{document}